\documentclass[12pt,preprint]{aastex}

\renewcommand {\deg}   {\mbox{$^\circ$}}

\newcommand   {\kms}   {\mbox{km\,s$^{-1}$}}
\renewcommand {\ga}    {\mbox{\rlap{\hbox{\lower5pt\hbox{$\sim$}}}\hbox{$>$}}}
\renewcommand {\la}    {\mbox{\rlap{\hbox{\lower5pt\hbox{$\sim$}}}\hbox{$<$}}}

\begin{document}



\def\kms {\hbox{km{\hskip0.1em}s$^{-1}$}} 
\def\msol{\hbox{$\hbox{M}_\odot$}}
\def\lsol{\hbox{$\hbox{L}_\odot$}}
\def\kms{km s$^{-1}$}
\def\Blos{B$_{\rm los}$}
\def\etal   {{\it et al. }}                     
\def\psec           {$.\negthinspace^{s}$}
\def\pasec          {$.\negthinspace^{\prime\prime}$}
\def\pdeg           {$.\kern-.25em ^{^\circ}$}
\def\degree{\ifmmode{^\circ} \else{$^\circ$}\fi}
\def\ee #1 {\times 10^{#1}}          
\def\ut #1 #2 { \, \textrm{#1}^{#2}} 
\def\u #1 { \, \textrm{#1}}          
\def\nH {n_\mathrm{H}}

\def\ddeg   {\hbox{$.\!\!^\circ$}}              
\def\deg    {$^{\circ}$}                        
\def\le     {$\leq$}                            
\def\sec    {$^{\rm s}$}                        
\def\msol   {\hbox{M$_\odot$}}                  
\def\i      {\hbox{\it I}}                      
\def\v      {\hbox{\it V}}                      
\def\dasec  {\hbox{$.\!\!^{\prime\prime}$}}     
\def\asec   {$^{\prime\prime}$}                 
\def\dasec  {\hbox{$.\!\!^{\prime\prime}$}}     
\def\dsec   {\hbox{$.\!\!^{\rm s}$}}            
\def\min    {$^{\rm m}$}                        
\def\hour   {$^{\rm h}$}                        
\def\amin   {$^{\prime}$}                       
\def\lsol{\, \hbox{$\hbox{L}_\odot$}}
\def\sec    {$^{\rm s}$}                        
\def\etal   {{\it et al. }}                     

\def\xbar   {\hbox{$\overline{\rm x}$}}         

\shorttitle{}
\shortauthors{}

\title{Signatures of an Encounter Between the G2\\
Cloud and a Jet from Sgr A*}
\author{Farhad Yusef-Zadeh$^1$ and Mark Wardle$^2$}
\affil{$^1$Department of Physics and Astronomy and Center for Interdisciplinary Research in Astronomy, 
Northwestern University, Evanston, IL 60208}
\affil{$^2$Department of Physics and Astronomy
and Research Center for Astronomy, Astrophysics \& Astrophotonics, Macquarie University, Sydney NSW 2109, Australia}


\begin{abstract} 
The recent discovery of the G2 cloud of dense, ionized gas on a trajectory toward Sgr A*, the black hole at the dynamical 
center of the Galaxy, offers a unique opportunity to observe an accretion event onto a massive black hole as well as to 
probe its immediate environment.  Simulations and models predict increased X-ray and radio variability resulting from 
increased accretion driven by drag on an atmosphere of hot, X-ray emitting gas surrounding Sgr A*.  Here, we present 
X-ray and radio light curves of the emission resulting from the potential encounter of the G2 cloud with a relativistic 
jet from Sgr A*. This interaction would violently shock a portion of the G2 cloud to temperatures $\sim 10^8$\,K 
resulting in bright X-ray emission from the dense, shocked gas as it adiabatically expands. The 2-10\, keV luminosity may 
reach $\sim$10 times the quiescent X-ray flux of Sgr A*. 
approximately $3\,\lsol$ is emitted above 10\,keV at the peak of 
the light curve, with significant softening of the spectrum occurring as the gas subsequently cools.  Observations with 
NuSTAR would therefore be able to confirm such an event as well as determine the cloud speed. At radio wavelengths, the 
associated synchrotron radio emission may reach levels of a few Jy. 
\end{abstract}



\section{Introduction}

Recent observations of the Galactic center at near-IR wavelengths have identified a dense, dusty gas cloud (dubbed ``G2'' by 
Gillessen et al 2012, 2013; see also Phifer et al. 2013) on a collision course with the compact radio source Sgr A*, the 
$4.3\times10^6$ \msol\, black hole located at the dynamical center of our Galaxy (e.g.\, Ghez \etal 2008; Gillessen \etal 
2009).  The cloud's mass is estimated to be $\sim$3 Earth masses with electron density $\sim2.6\times10^5$ 
cm$^{-3}$, and dust and gas temperatures of $\sim$600 K, and $\sim10^4$ K, respectively.  G2 is on a highly eccentric orbit 
(e$\sim0.97$) that is coplanar with the clockwise stellar disk (Paumard \etal 2006; Lu \etal 2009; Bartko \etal 2009) and is 
expected to reach its pericenter distance of $\sim2000$ Schwarzschild radii (R$_S$) from Sgr A* in mid September, 2013 
(Gillessen et al. 2013) or March 2014 (Phifer et al. 2013).

The origin and nature of the G2 cloud is unclear - models include a pressure-confined compact gas cloud (Gillessen \etal 
2012), an extended spherical shell created by colliding stellar winds (Burkert \etal 2012; Schartmann \etal 2012), a 
photo-evaporative wind from a proto-planetary disk surrounding a low-mass star (Murray-Clay \& Loeb 2012), a shell from a nova 
outburst (Meyer \& Meyer-Hofmeister 2012), or an outflow from a T Tauri star (Scoville \& Burkert 2013), in which case the 
observed emission arises from the ionized surface of a more extensive cloud.  The change in the kinematic structure of the 
emission, however, is consistent with simple tidal stretching, and the mass of ionized gas inferred from the Br\,$\gamma$ 
emission has not changed significantly over 4 years of monitoring, casting doubt on some models.

G2 is being tidally sheared (Gillesen \etal 2013) as it approaches pericenter, and will eventually be disrupted.  Shredding of 
the cloud due to drag on the hot, tenuous medium thought to be responsible for the X-ray emission within 0.5$''$ of Sgr A* may 
hasten this and lead to the accretion of a substantial fraction of G2 on to the black hole (e.g. Gillesen et al 2012; Burkert 
\etal 2012; Schartman et al 2012; but see also Anninis et al 2012 and Saitoh et al 2012).  This interaction is also expected 
to produce a weak bow shock in the pre-existing hot quasi-vitalized gas settling in Sgr A*.  Electrons accelerated in the 
shock may increase the radio flux associated with Sgr A* by an order of magnitude (Narayan \etal 2012; Sadowski \etal 2013).  
In addition, Bartos \etal (2013) have discussed the accretion of some of G2 by a population of stellar black holes and neutron 
stars that may be residing close to the dynamical center of the Galaxy because of mass segregation in the inner nuclear 
cluster centered on Sgr A*.

Here, we point out that G2 may also interact with a jet emerging from Sgr A* as has been tentatively detected by Yusef-Zadeh 
at al 2012, and previously proposed to explain the SED and the variability in the emission from Sgr A* (e.g.\ Markoff \etal 
2007, Falcke et al.\ 2009).



The broad-band spectrum of the quiescent emission from Sgr A* has been modeled in terms of emission from the base of a 
relativistic jet, and Very Long Baseline Array measurements suggest that the position angle (PA) of the outflowing 
material is $\sim90^0$ (Markoff \etal 2007).  Short time scale variability of Sgr A* and the time delay between the peaks of 
radio emission at different frequencies has also been interpreted in the the context of a jet or outflow from Sgr A* 
(Yusef-Zadeh \etal 2006; Maitra \etal 2009).  These measurements suggest a mass outflow rate $\la\,2\times10^{-8}$ \msol\, 
yr$^{-1}$.

On larger, but sub-parsec scales, a number of observations indicate a collimated outflow or a jet from Sgr A*.  A chain of 
radio blobs links Sgr A* to the``minicavity'' at 8.4 GHz (Yusef-Zadeh, Morris \& Ekers 1990; Wardle \& Yusef-Zadeh 1992).  The 
minicavity, roughly at PA$\sim60^0$ from Sgr A*, shows enhanced FeIII line emission, a high electron temperature and is 
kinematically disturbed (Eckart \etal 1990; Lutz \etal  1993). Infrared observations have also discovered two dust features 
with cometary morphology along the line between Sgr A* and the mini-cavity (Muzic \etal 2010), their orientation is 
suggestive of an outflow from young massive stars or from Sgr A* (Lutz \etal  1993; Muzic \etal  2009).

Radio observations of Sgr A* have provided a tantalizing detection of a jet-like linear feature emanating 
symmetrically from Sgr A* on a scale of $\sim3$\,pc (Yusef-Zadeh \etal 2012).  A number of radio continuum 
features with X-ray and FeII line counterparts, and linearly polarized features are also detected along the 
axis of this 3\,pc jet feature.  The feature is projected at an angle that appears to be parallel to the 
angular momentum axis of the clockwise stellar disk and the orbital plane of G2.  The axis of the jet from Sgr 
A* is estimated to be at PA$\sim60^0$. Model fitting of 
the polarization of the near-IR emission from Sgr A* infers  
the spin axis of Sgr A* to be  similar to the PA of the jet (Zamaninasab \etal 2011).  
Based on the interaction of the jet and the minicavity, 
the mass outflow rate in the jet is 
estimated to be 10$^{-6}$ \msol\, yr$^{-1}$, with a Lorentz factor $\gamma\sim$3.


This motivates us to investigate potential signatures of the interaction of a jet emanating from Sgr A* and the G2 cloud.  
The collimated appearance of the jet on parsec scales suggest that the jet close to Sgr A* must be narrow with a 
diameter of $d_j\sim3000\,R_S$.
Figure 1 shows a schematic diagram of the geometry of the jet and G2 as they interact with each other.  In this scenario, the 
G2 cloud is assumed to be tidally stretched as it sweeps across the 
barrel of the jet.  Assuming that the ram pressure in the jet greatly exceeds that associated with the orbital motion of G2 
(i.e. the outflow rate $\ge 10^{-9}$ \msol\, yr$^{-1}$), the jet is not significantly deflected by the interaction, but 
instead acts as a solid obstacle.  As the orbital speed of G2 is hundreds of times its internal sound speed, a bow shock 
standing off from the jet forms within G2 as it streams past, as sketched in Figure 1.  The hot ($T\sim10^8$\,K) and dense 
($\nH\sim10^6\ut cm -3 $) shocked gas emits X-rays, but cools primarily adiabatic expansion parallel to the jet axis into the 
under pressured external medium.  Relativistic electrons are accelerated in the bow shock, producing synchrotron emission.  


\section{Estimates and light curves}

Before computing model light curves for this interaction, 
 we first adopt nominal values of the critical physical parameters and 
estimate the strength of the emission expected in X-rays and  radio. 
Before being tidally sheared G2 had a diameter $\sim 300\,R_S$ and number density of hydrogen nuclei $3\ee5 \ut cm -3 $, 
corresponding to a mass $10^{-5}$\,\msol\ (Gillesen et al 2011, 2012).  G2 will be tidally sheared along its orbit and 
compressed perpendicular to it (e.g. Saitoh \etal 2012).  Here we assume that at the point of interaction with the jet, the 
cloud is stretched to a length $L_g = 10^4 R_S$, while its transverse dimension is $L_j=10^3 R_S$.  This combination of 
dimensions implies that the cloud density is very close to its initial density, so we simply adopt $n_g = 3\ee 5 \ut cm -3 $ 
as the hydrogen  density.  
The temperature in the cloud is maintained close to $10^4$\,K by efficient cooling (Saitoh \etal 2012).  
The sound speed is therefore $\approx 10$\kms.  At pericenter the orbital velocity is very close to the escape velocity, ie 
6\,400 \kms.  We choose a lower nominal shock speed, $v_g=3\,000$\,\kms, partly because interaction may occur prior to or 
subsequent to periastron, and because the cloud velocity may not be perpendicular to the jet.  The final quantity that we 
require is the diameter of the jet, which we set at $d_j = 300\,R_S$.  This implies an opening angle of about 8$^\circ$, 
consistent with estimates based on tentative detection at radio wavelengths (Yusef-Zadeh et al 2012).


Now we consider the emission from the immediate post shock gas, which has a temperature $T = 3\mu v_g^2/16k = 1.25\ee 8 $\,K, 
where k is Boltzmann's  constant and $\mu$ is the mean particle mass,  
number density $n_H=4n_g=1.2\ee 6 \ut cm -3 $, and emits X-rays with a radiative power per unit volume $\Lambda = \nH^2 f(T)$
where we  approximate the results of B\"ohringer \& Hensler (1989) for gas of twice solar metallicity by\\ 
$f(T) \approx 1.5\ee -23 ((T/4\ee 6 )^{-2.75} + (T/4\ee 6 )^{0.8} )\, \u erg\, \ut s -1 \ut cm 3 $
when $2\times10^6<T<10^7 $\, K\\
 and $f(T) \approx 1.2\ee -23 ((T/1.5\ee 7 )^{-1.4} + (T/4\ee 7 )^{0.5} )\, 
\u erg\, \ut s -1 \ut cm 3 $ when  T$> 10^7 $\, K. 
The shock interaction creates thermal energy at a rate \begin{equation}
    L_\mathrm{shock} = \textstyle{\frac{1}{2}} \rho_g v_g^3 L_j d_j \approx 1200 \; \lsol \,.
    \label{eqn:Lshock}
\end{equation}
over a period
\begin{equation}
    t_\mathrm{int} = \frac{L_g}{v_g} = 1.3 \u yr .
    \label{eqn:tint}
\end{equation}
The radiative timescale for the shocked gas, 
\begin{equation}
    t_\mathrm{rad} = \frac{5}{2}\, \frac{kT}{\rho f(T)} \approx 72  \u yr
    \label{eqn:trad}
\end{equation}
is long compared to its timescale for expansion,
\begin{equation}
  t_\mathrm{exp} = \frac{L_j}{c_s} = 0.24 \u yr \,.
  \label{eqn:texp} \end{equation} so the gas cools by adiabatic expansion.  Thus the peak luminosity is approximately the 
power radiated by the hot gas generated by the shock over an expansion time scale.  The thermal energy in this gas is 
approximately $L_\mathrm{shock} t_\mathrm{exp}$, yielding an estimate \begin{equation}
    L_X   \approx  \frac{L_\mathrm{shock} t_\mathrm{exp}}{t_\mathrm{rad}}\, \approx 4.7 \, \lsol.
    \label{eqn:Lx}
\end{equation}
 This 
luminosity exceeds the $\sim 2.4\ee 33 \u erg \ut s -1 $ quiescent X-ray luminosity of 
Sgr A*  based on  Chandra observations (Baganoff et al. 2003),
so should be  detectable with the  Chandra, XMM, and Swift observatories.


We adopt an analogous approach to estimating the radio flux arising from synchrotron emission from relativistic electrons 
accelerated by the shock.  We assume that the power that the shock wave deposits into relativistic electrons is $\epsilon 
L_\mathrm{shock}$, with a nominal efficiency $\epsilon=0.01$\footnote{This is distinct from the fraction $\eta=0.05$ of 
shocked electrons that are assumed to be accelerated by Narayan et al (2012) and Sadowski et al (2013), equivalent to 
$\epsilon\approx 2$  and 0.4, respectively.}, and that the electrons are produced with an $E^{-2}$ spectrum.  Here the 
relevant radiative cooling time is the synchrotron loss time scale for the electrons that dominate the emission at, say, 
$\nu=1.4$\,GHz, so we also need estimate the post shock field strength, which we parametrize as  $B^2 = 8\pi \epsilon_B \rho_g 
v_g^2$.  In supernova remnant (SNR)  shocks the magnetic field is amplified by cosmic ray acceleration to levels $\epsilon_B \sim 
10^{-2}$--$10^{-3}$ (Vink 2012; Helder et al 2012; Ellisen et al 2012), so we adopt $\epsilon_B=0.01$, yielding $B \approx 
0.12$\, G.  These electrons have energy $E_\nu \approx 26$\,MeV, with loss time \begin{equation}
    t_\mathrm{synch} \approx 20\, \u yr \,.
    \label{eqn:tsynch}
\end{equation}
Again, the radiative loss time is long, so that we can use a similar expression as for the X-rays to estimate the synchrotron luminosity, and then use $L_\mathrm{synch} \approx 4 \pi d^2 \, \nu S_\nu$ to find the flux, yielding
\begin{equation}
    S_\nu  \approx \frac{1}{4 \pi d^2 \nu} \, \frac{\varepsilon L_\mathrm{shock} t_\mathrm{exp}}{t_\mathrm{synch}}  \approx
    5.1\, \u Jy.
    \label{eqn:Snu}
\end{equation}
which significantly exceeds the $\sim 1$\,Jy quiescent flux from Sgr A* at 1.4\,GHz.


To compute approximate X-ray light curves, we characterize the shocked gas by its cross-sectional area $A$, density, 
temperature, and velocity $v$ perpendicular to $A$.  These quantities all evolve as the gas moves downstream.  The 
cross-section expands at a rate $dA/dt = L_j\, c_s$ where $c_s$ is the adiabatic sound speed, and the associated drop in 
density and increase in velocity are found by conservation of mass and momentum and noting that the expansion is adiabatic, so 
that $\rho A v$, $(\rho v^2 + P) A$, and $P/\rho^{5/3}$ are constant.  The resulting profile allows us to compute the X-ray 
emissivity by integrating the volume emissivity of the 2-10\, keV X-ray emission, ie $\Lambda \left[\exp(-2\u keV /kT) - 
\exp(-10\u keV /kT)\right]$, as the gas moves downstream.  We have not accounted for extinction but this is not expected to be 
severe as the spectrum is hard ($kT\sim 10$\,keV).

The synchrotron light curves are computed based on the assumption that the field and electrons are advected along with the 
postshock gas.  The magnetic field must be tangled if it is significantly larger than the preshock field, so it's strength 
scales as $\rho^{2/3}$.  Meanwhile, the electrons' individual energies scale as $\rho^{1/3}$ while preserving their spectral 
index: as a result the electron and magnetic energy densities both scale as $\rho^{4/3}$ and  the synchrotron emissivity 
scales as $\rho^{7/3}$.

The resulting X-ray and 1.4 GHz light curves for our fiducial parameters are shown in Figure 2, respectively For our nominal 
parameters (solid curves), the X-ray light curve (upper panel) begins sharply once shocked gas starts to be created by the 
interaction of G2 with the jet.  The luminosity rises rapidly as the mass of shocked gas increases, but the rate of increase 
in $L_X$ declines once adiabatic expansion starts to cool the oldest gas. During this phase, the net change in the shocked gas 
is the addition of successively older and cooler layers of gas at the extreme downstream end of the shocked gas, and the 
change in luminosity corresponds to the emission from this oldest layer.  The rise in luminosity terminates once the 
interaction ends and the cooling gas is no longer being supplemented by newly shocked material.  As the luminosity is 
dominated by this material, the initial decline is sharp, but then tails off as the shocked gas slowly cools off.  The peak 
luminosity is consistent with the rough estimate given by eq (5); the extreme sharpness of the initial decline from the peak 
reflects the artificially sharp truncation of shock heating in our treatment.  The 1.4 GHz light curve behave similarly, as 
shown in the lower panel of Figure 2.
Note that the peak flux density at 1.4 GHz is few times lower than the estimate in eq (7) because significant emission occurs 
at higher frequencies. 
Also plotted are the light curves for different choices of shock speed.  At lower speeds the X-ray luminosity and synchrotron emission drop sharply, because the shock luminosity depends quadratically on the cloud speed $v_g$.  At higher speeds, the radio flux increases similarly, but the X-ray flux does not because the gas temperature is so high that much of the X-ray emission emerges  at energies exceeding 10\,keV.

We illustrate this in the top panel of Figure  3 by comparing the luminosities in different X-ray bands between 5 and 80 keV for our 
nominal shock speed of 3000 km/s.  The peak luminosities between 5-10 and 20-40 keV are similar because the shocked gas initially 
has $kT \approx 10$\,keV.  Once the interaction between G2 and the jet ceases, the spectrum becomes softer as the shocked gas cools 
by adiabatic expansion.  The lower panel shows the effect of increasing $v_g$ to 5\,000 \kms.  Now the gas is shocked to $kT \approx 
25$\,keV, with a corresponding dramatic increase in the power radiated between 20-40 and 40-80 keV relative to the lower energy 
bands.  We conclude that observations with NuSTAR would be ideally suited to determine the gas temperature and determine the shock 
speed.

\section{Discussion and Conclusion}

We have shown that if the G2 cloud encounters a jet from Sgr A* during its passage around pericenter then it is likely to 
produce detectable X-ray emission. The X-ray luminosity is far larger than expected from the drag interaction of G2 with a hot 
medium centered on Sgr A* because of the violence with which the cloud encounters the jet, which shocks a fraction of the 
cloud to high temperatures and high density.  The greatest uncertainties in estimating the luminosity are the density and 
geometry of the incoming cloud, which may be significantly compressed by the tidal effects of Sgr A* (Saitoh et al 2012). 
Close to pericenter G2 will be strongly compressed in the direction perpendicular to the orbital plane (Saitoh et al 2012), 
and the density will increase by orders of magnitude. This increases the emission from the shocked gas, which is proportional 
to $n_g^2\, \rm L_j$, but decreases the probability of interaction with the jet. The decays in the curve due to adiabatic 
expansion is also unique to this model, reflecting the fact that the shocked cloud material is over pressured.  Observations 
using NuSTAR would provide a strong test of this scenario, as the high temperature and subsequent expansion of the shocked gas 
produces detectable spectral evolution in the X-ray light curves between 5 and 80\,keV.
We also predict a detectable radio flux, based on the assumption that 1\% of the kinetic luminosity is deposited into 
relativistic electrons and that the magnetic field is amplified to a level similar to that inferred for supernova remnants.  


The greatest uncertainty is whether a jet from Sgr A* is sufficiently aligned with the orbital plane of G2 for an 
interaction to even occur.  Figure 4 shows the range of jet inclinations that lie within 5\deg\ of the orbital 
plane as a function of jet PA.  
We have plotted the inclination's cosine so that solid angle is directly 
proportional to area, allowing assessment of probabilities. A randomly-oriented jet has probability $\approx 
0.087$ of lying within 5\deg\ of the plane, precisely the fraction of the plot area lying within the target shaded 
region.  The blue rectangles indicate the ranges of jet orientation previously inferred either from Sgr A* 
emission models or from parsec-scale jet-like features identified in radio and X-rays: (i) combined disk/jet 
models of the emission at 7\,mm detected using VLBA (MO7 -- Markoff et al.\ 2007), (ii) the normal to the 
accretion disk orientation inferred from modeling the sub-mm VLBI observations of Sgr A* (B11 -- Broderick et al.\ 
2011), (iii) polarization of NIR flaring (Z11-Zamaninasab et al.\ 2011), (iv) a 0.5\,pc jet-like feature 
identified in X-rays (M08 -- Muno et al.\ 2008), and (v) a parsec-scale radio jet (YZ12 -- Yusef-Zadeh et al 
2012).  In the latter two cases, the inclination ranges are very uncertain. We adopt $\pm45$\deg from the plane of 
the sky for the X-ray jet. For the radio jet, the PA$\approx238$\deg\ side of the jet is directed between 
$30$--$60$\deg\ in front of the plane of the sky so that it can collide with material to form the minicavity 
(Yusef-Zadeh et al.\, 2012). The uncertainty in both jet orientation and the orbital plane make a reliable 
estimate impossible, but given the tendency of the inferred jets to follow the orbit of G2, odds of 1 in 3 or 4 
are not implausible.

The center  and lower panels of Figure 4 show the shock speed and epoch of a potential interaction as a function of jet PA.  The shock 
speed is equated to the azimuthal component of G2's orbital velocity as this will be normal to the jet.  As we expect significant 
emission only for shock speeds exceeding 2000 km/s (unless the density of G2 is significantly enhanced by tidal effects), the relevant 
range of PA and epoch is $+100$ to $-100$\deg and early 2013 - mid-2014 for the Gillesen et al (2013) orbit, and $-140$\deg to 
$+70$\deg and mid-2013 - end 2014 for the Phifer et al.\ (2013) orbit.  In either case we expect appreciable emission if the cloud 
collides with B11/Z11/Y12 jets at PA$\sim60$\deg and the M07, M08 jets at PA$\sim 85$\deg.

In summary, we have described a model in which the G2 cloud runs into a relativistic jet from Sgr A* and have predicted X-ray, 
and radio light curves that are distinct from those produced from the collision of G2 with the hot atmosphere of Sgr A*.  
Monitoring the emission from Sgr A* at hard X-rays using NuStar may be particularly useful in testing the proposed model.


\acknowledgments 
This work is partially supported by grants AST-0807400 from the NSF and DP0986386 from the Australian Research Council.
\newcommand\refitem{\bibitem[]{}}
{}

\begin{figure}
\center
\includegraphics[scale=0.8,angle=0]{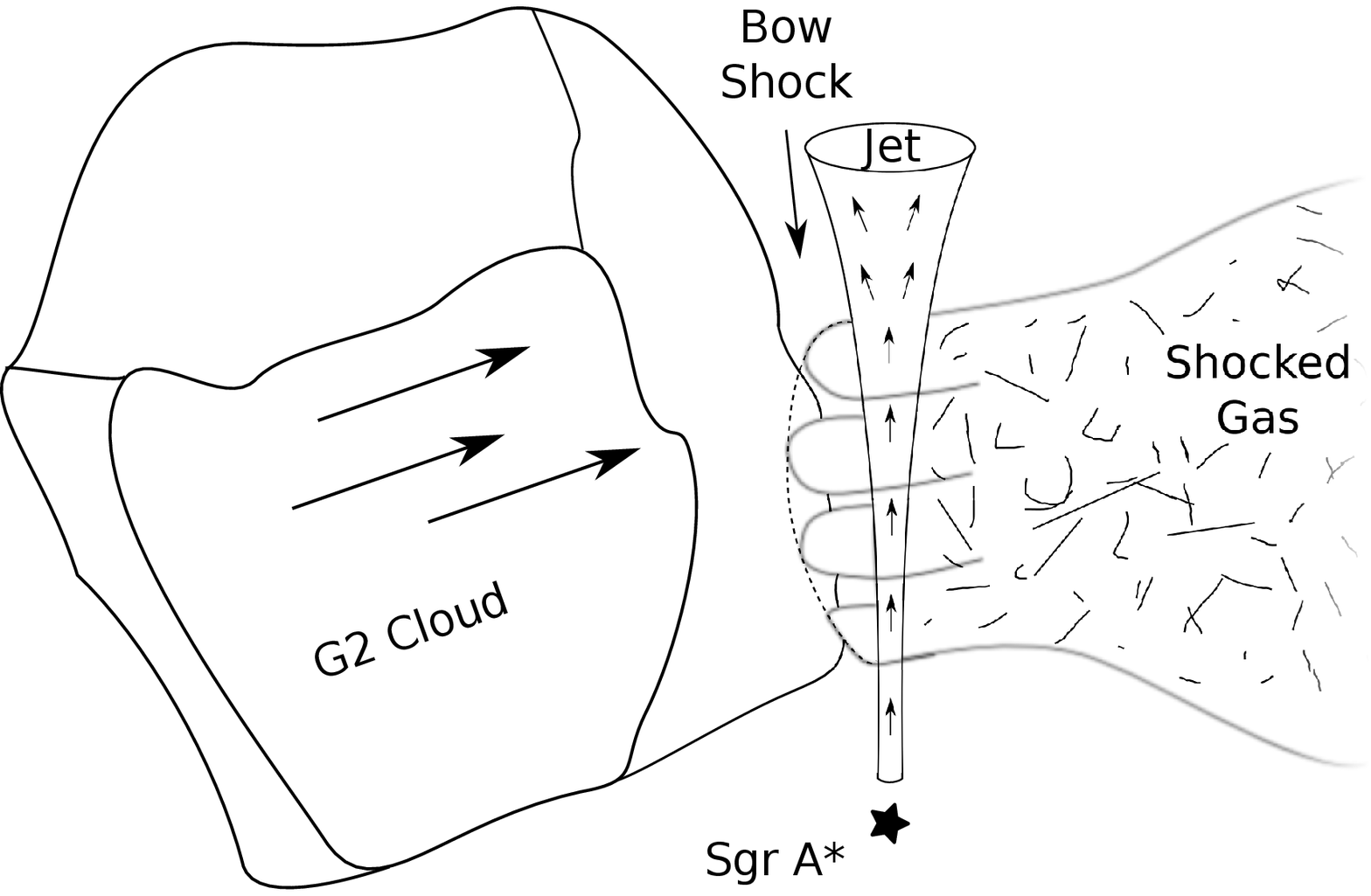}\\
\caption{
A  Schematic diagram of a model in which the G2 cloud runs into a relativistic jet emanating from 
Sgr A*. 
}
\end{figure}  

\begin{figure}
\center
\includegraphics[scale=0.8,angle=0]{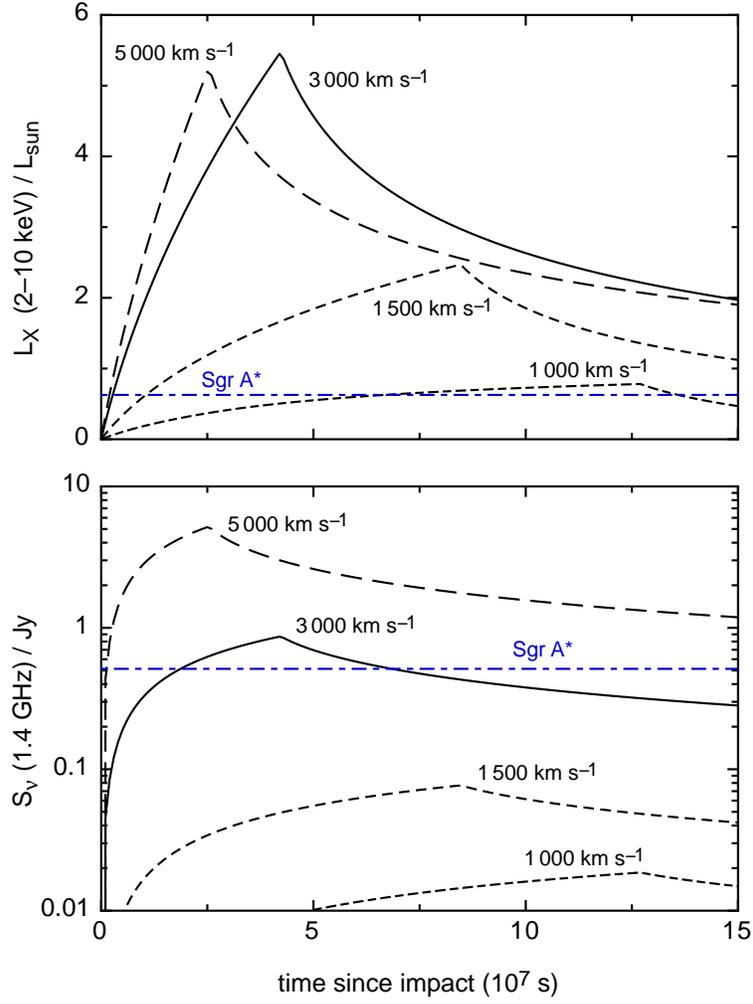}\\
\caption{
X-ray  (upper panel) and radio light curves 
lower panel) arising from the interaction of the cloud with a jet 
from Sgr A*, for different values of the G2 cloud velocity normal to the jet axis.
The blue dot-dashed line in each panel indicates the level of the quasi-steady 
emission from Sgr A* (Zhao et al 2001; Baganoff et al. 2003).
}
\end{figure}  

\begin{figure} 
\center 
\includegraphics[scale=0.8,angle=0]{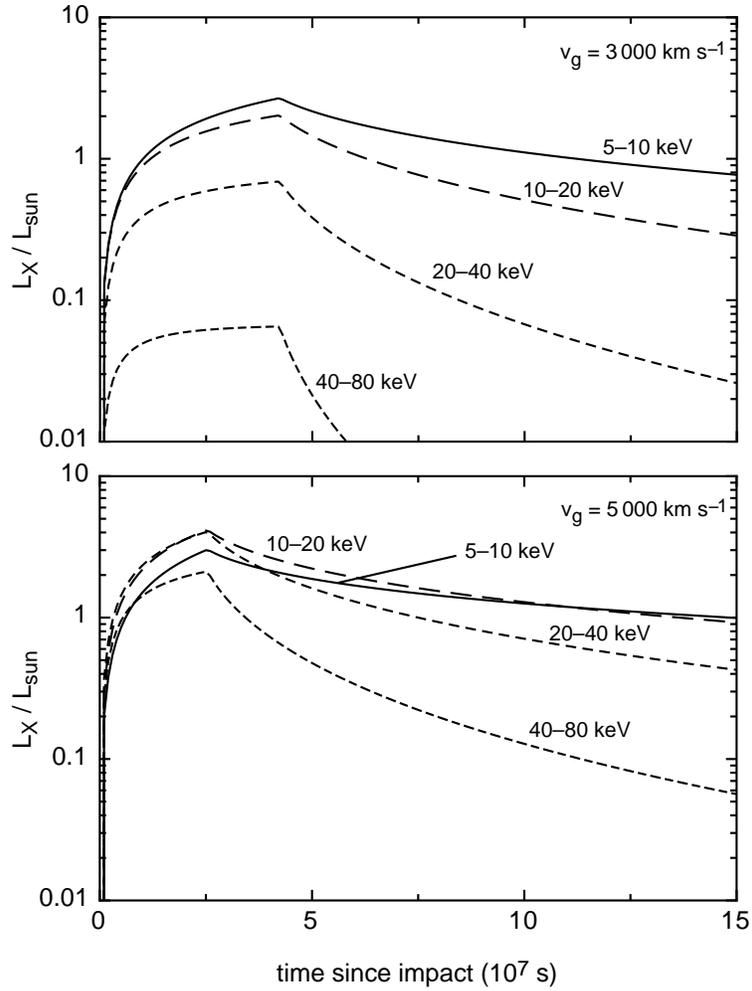}
\caption{ Light curves in different X-ray bands 
arising from the collision of the G2 cloud with a jet from Sgr A* for cloud velocities of 3000 \kms\, (top panel) 
and 5000 \kms\, (lower panel).} 
\end{figure}

\begin{figure}
\center
\includegraphics[scale=0.8,angle=0]{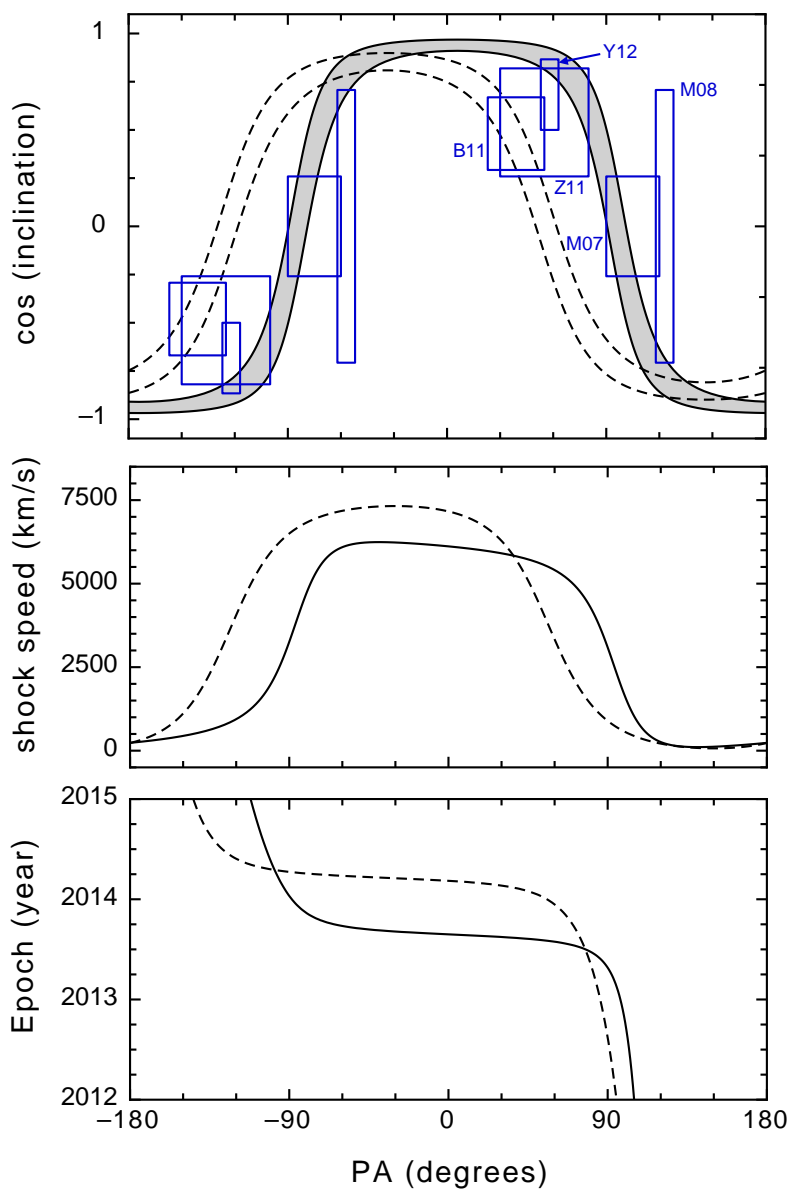}
\caption{
The upper panel shows the possible orientations of a Sgr A* jet directed within 5\deg\ of the G2 orbits inferred 
by Gillessen et al. (2013) (shaded) and Phifer et al. (2013) (dashed lines). PA is measured E of N and inclination 
is measured relative to the line of sight towards the observer. The rectangular regions indicate possible jet 
orientations based on Sgr A* emission models or jet-like features identified in radio and X-rays (see text). 
The  center and lower panels show  the PA-dependence of the shock speed and interaction epoch, respectively, 
for the G2 orbits determined by  Gillessen et 
al.\ (2013; solid lines) and Phifer et al. (2013; dashed lines).
}
\end{figure}

\end{document}